\documentstyle[aps,prl,psfig,floats]{revtex}

\begin{document}
\draft
\title{An Optical Lattice of Ring Traps}
\author{Philippe Verkerk and Daniel Hennequin}
\address{Laboratoire de Physique des Lasers, Atomes et Mol\'{e}cules, UMR\\
CNRS,\\
Centre d'Etudes et de Recherches Lasers et Applications, Universit\'{e} des\\
Sciences et Technologies de Lille, F-59655 Villeneuve d'Ascq cedex - France}
\date{\today }
\twocolumn[\hsize\textwidth\columnwidth\hsize\csname @twocolumnfalse\endcsname
\maketitle

\begin{abstract}
A new geometry of optical lattice is proposed, namely a lattice made of a 1D
stack of ring traps. It is obtained though the interference pattern of two
counterpropagating beams: one of the beam is a standard gaussian beam, while
the other one is a hollow beam obtained through a setup with two conical
lenses. The resulting lattice is shown to have a high filling rate and a
good confinement, so that it could be loaded directly from a MOT with
applications in the domain of quantum computing, or with a Bose-Einstein
condensate, which would have in this case a 1D ring geometry.
\end{abstract}


\vskip2pc] 

\renewcommand\floatpagefraction{.9} 
\renewcommand\topfraction{.9} 
\renewcommand\bottomfraction{.9} 
\renewcommand\textfraction{.1} 

Optical lattices are obtained by dropping atoms in the interference pattern
of coherent laser beams. Atoms undergo a force whose potential is
proportionnal to the light intensity, so that cold atoms are located in
confined sites, corresponding to the maximal (zero) intensity for bright
(dark) lattices. By varying the shape of the interference pattern, atoms can
be manipulated with an extreme precision and a relative ease: because of
these possibilities, optical lattices have recently attracted increasing
interest in various domains. Indeed, lattices may be loaded with atoms
cooled in a magneto-optical trap (MOT), or with a Bose-Einstein condensate
(BEC). Several spectacular results have been obtained recently with the
latter configuration, as e.g. the realization of a BEC coherent accelerator
and multiphoton beam-splitter\cite{Accelerator}, and the experimental
demonstration of the quantum phase transition from a superfluid to a Mott
insulator\cite{MottBEC}. It is also possible to study more fundamental
properties of the BEC, as e.g. their free expansion in a low-dimensionnal
lattice\cite{Expansion}, or their collective excitations\cite{Excitations}.
Another attractive research field concerns low-dimensional BEC. In
particular, recent experiments have demonstrated the formation of bright
solitons in a 1D BEC with attractive interactions\cite{1DBEC}. These
experiments have been realized with simple waveguides, and thus a linear
geometry of the BEC. Optical lattices could enable the use of more complex
geometries, leading in particular to ring traps, in which BECs should
exhibit new properties\cite{RingBEC}.

Optical lattices are also a key system in the realization of quantum
computers with neutral atoms\cite{QComp2}. The qubits are the two-level
atoms trapped in the lattice sites, and the interaction occurs through a
cold collision between two atoms in neighboring sites. This requires to
produce a lattice with at least one atom by site, a good confinement of the
atoms in each site and a decoherence as low as possible. When the lattice is
loaded from a MOT, the first condition is not fulfilled, simply because of
the relative weak atomic density in the MOT (typically 10$^{11}$ atoms/cm$%
^{3}$), as compared to the relative high density of sites in the lattice
(typically 10$^{13}$ sites/cm$^{3}$ in a 3D square lattice). A possibility
is to limit the number of sites by using a 1D lattice, but in this case, the
other conditions are not fulfilled. Indeed, if a bright lattice is used, the
interactions of the trapped atoms with the light field is maximal, and the
coupling between atoms and light modifies, through different physical
processes, the light field, and disturb the trapping and cooling mechanisms%
\cite{BrightLattice}. Moreover, when such a lattice is near resonance, it
induces decoherence, whereas at large detuning, the confinement of atoms,
linked to the wavelength of light, decreases. In the same way, when a dark
lattice is used, the atoms are not trapped, except in the particular
situation of a lattice of toroidal trap. But in the proposed configurations,
the radial confinement is low, of the order of the beam waist\cite{Cavity}.
Until today, the only lattices able to strongly confine atoms with
fluorescence rates of the order of 1 s, were the 3D dark lattices\cite%
{DarkLattice}. Therefore, a solution for a high filling rate is to increase
the density of the atoms dropped in a 3D dark lattice, by using a BEC\cite%
{MottBEC}. However, note that in this case, the BEC is not use for its
particular coherent properties, but only because it makes possible the
access to higher atomic densities.

We propose here a new lattice geometry, where each site is a ring trap,
while the site density is typically 10$^{8}$ sites/cm$^{3}.$ Thanks to its
properties, in particular the high confinement of the sites, and to its
great adaptability, many applications of such a lattice can be considered.
In particular, it can be loaded with a BEC, and thus could produce a 1D ring
BEC, but thanks to the low density of sites, it can also be loaded directly
with a MOT. The paper is organized as follows: we first discuss the
principle of the lattice of ring traps, then describe the experimental
realization, and finally show basic results concerning atoms loaded in the
lattice.

Each individual trap is a 3D dark ring, and the lattice is a 1D stack of
such rings. Thus, the global shape of the potential is a bright full
cylinder with a pile of ring wells inside. Such potentials can easily be
obtained by making the interference of a standard gaussian beam with a
counterpropagating hollow beam, as illustrated schematically in fig. \ref%
{fig:principle}a. Both beams have the same blue detuned frequency, so that
the trapping sites correspond to the zero intensity sites. They propagate
along the $z$ vertical axis with the same polarization, and the hollow beam
has an inverse gaussian distribution along the radial direction $r$, with a
zero intensity in its center. The interference pattern is shown on fig. \ref%
{fig:principle}b through the longitudinal evolution of the transverse
distribution. When the two beams are out of phase (on the figure, in $%
z=\lambda /4+n\lambda /2$ with $n$ integer and $\lambda $ the wavelength),
two zeros of intensity appear symmetrically on each side of the center.
Because of the cylindrical symmetry, these zeros correspond to a unique ring
along the azimutal direction. On the contrary, when the two beams have the
same phase (in $z=n\lambda /2$), the intensity profile reaches its maximum.
This interference pattern results in a potential $U$ which is, in the limit
of weak saturation and large detuning, proportional to the light intensity $%
I $:%
\begin{equation}
U=\frac{\hbar }{8}\frac{\Gamma ^{2}}{\Delta }\frac{I}{I_{S}}
\end{equation}%
where $\Delta $ is the detuning, $I_{S}$ the saturation intensity and $%
\Gamma $ the width of the atomic transition. This potential is a stack of
rings separated by $\lambda /2$. Each ring has a radius $r$ of the order of
the waist $w_{0}$ of the gaussian beam, and a thickness $\Delta r$ also of
the order of $w_{0}$. If $z$ is chosen as the vertical axis, the cylindrical
symmetry of the potential is not broken by gravity.

\begin{figure}[tph]
\centerline{\psfig{figure=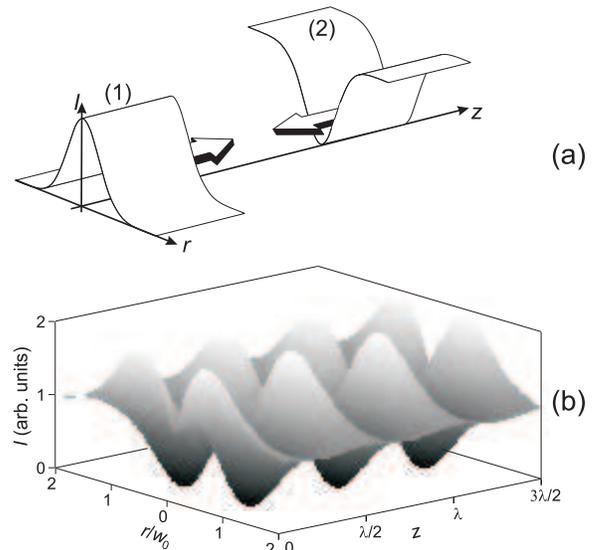,width=3in}}
\caption{Principle of the ring potential. In (a), schematic representation
of the beam configuration, through the propagation of their intensity along
the $z$ axis: the full (1) and the hollow (2) beams are counterpropagating.
In (b), the resulting interference pattern. The transverse profile of the
intensity $I$ versus the radius $r$, is plotted versus the propagating
direction $z$.}
\label{fig:principle}
\end{figure}

In the context described above, two main properties of the lattice must be
optimized: (i) the filling rate, and (ii) the confinement in the $z$ and $r$
directions. As the number of sites is fixed, the former is only linked to
the number of atoms loaded in the lattice.. This number is of the order of
the lattice volume, i.e. $\pi r^{2}L$, where $L$ is the fixed length of the
lattice, corresponding to the diameter of the atomic cloud used to load it.
Thus in order to increase the filling rate, $r$ must be chosen as large as
possible. On the other hand, the confinement following the $r$ direction is
linked to the radial thickness $\Delta r$ of the rings, which accordingly
must be chosen as small as possible. Therefore, the simple example given in
Fig. \ref{fig:principle}, appears not to be optimal for our lattice, as both 
$r$ and $\Delta r$ are proportional to $w_{0}$, while they should on the
contrary be independent, in order to be adjusted separately. Moreover, an
inverse gaussian beam has the disadvantage to concentrate most of the energy
in the sides of the beam, where it is not useful. To summarize, a good
configuration should concentrate most of the light intensity in the vicinity
of $r$, and make $r$ and $\Delta r$ independent.

\begin{figure}[tph]
\centerline{\psfig{figure=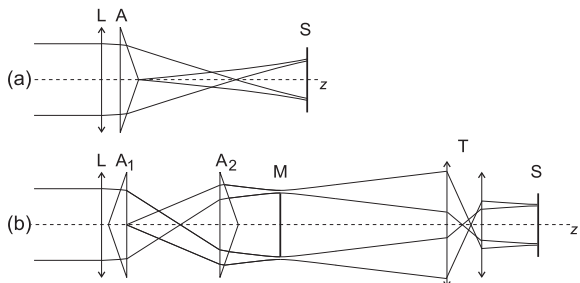,width=3in}}
\caption{Illustration of the beam propagation after one conical lens in (a),
and two conical lens in (b). In (a), the beam crosses the lens L and the
conical lens A. On the screen S, located beyong the Bessel zone, the
transverve profile of the beam is a series of concentric rings. In (b), the
beam crosses the lens L, then the two conical lenses A$_{1}$ and A$_{2}$.
The mask M eliminates the diffracting rings, and the telescope T reduces the
beam size. On the screen S, the transverve profile of the beam is a
collimated ring.}
\label{fig:axicon}
\end{figure}

\begin{figure}[tph]
\centerline{\psfig{figure=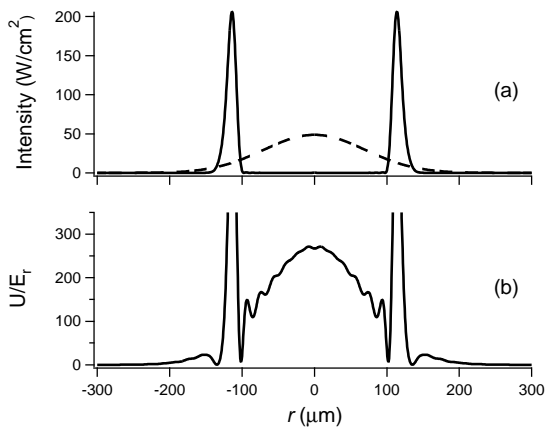,width=3in}}
\caption{In (a), theoretical transverse profile of the gaussian (dashed) and
hollow (full) beams used in the experiment. In (b), the resulting transverse
potential at the bottom of the well. Parameters are those used in the
experiments. }
\label{fig:pot}
\end{figure}

To obtain these properties, we use a hollow beam produced by a conical lens,
as described in \cite{Axicon}. The conical lens is an axicon, i.e. an
optical system that produces a line focus rather than a point focus from
incident collimated beam\cite{Axicon54}. Axicons are extensively used to
produce Bessel-Gauss beams\cite{BesselGauss} or annular beams\cite%
{Axicon78,Annular}. The standard construction to generate an annular hollow
beam is a doublet made by a converging lens and a conical lens, as
illustrated in fig. \ref{fig:axicon}a: the incident gaussian beam is
transformed in a ring, as each part of the beam is deviated towards the
optical axis $z$. In fact, because of the diffraction on the lens vertex,
the resulting beam exhibits, inside the main ring, secondary rings of lower
intensity \cite{Axicon}. However, these rings can be easily removed with a
mask, leading to a resulting annular beam with typically 80\% of the
intensity of the initial gaussian beam distributed in the unique ring (Fig. %
\ref{fig:pot}a). A second conical lens is used to collimate the radius $r$
of the ring, so that after this second lens, $r$ becomes constant with $z$
(Fig. \ref{fig:axicon}b). Note that the beam is not really collimated, as
its waist, which is now the thickness of the ring, evolves following the
usual rules of propagation. In particular, the lens L in Fig. \ref%
{fig:axicon}b produces a beam with a minimum thickness in M, where the mask
used to suppress the secondary rings is located. The resulting hollow beam
has a radius $r$ essentially dependent on the distance between the two
conical lenses, while its thickness $\Delta r$ depends on the focal length
of L. Thus $r$ and $\Delta r$ are adjustable independently.

\begin{figure}[tph]
\centerline{\psfig{figure=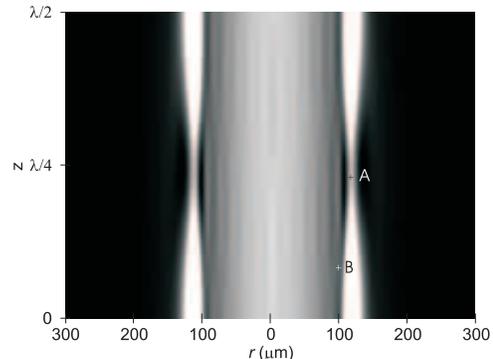,width=2.5in}}
\caption{2D representation of the potential as a function of the radius $r$
and the longitudinal coordinate $z$. Parameters are those of Fig.  \protect
\ref{fig:pot}. Note that the two scales are different. Dark corresponds to a
zero potential. Signification of points A and B is given in the text.}
\label{fig:pot2D}
\end{figure}

The potential is obtained by focusing the gaussian beam and the
counterpropagating hollow beam at the same point, so that the wave surfaces
are planes perpendicular to the propagation axis. The resulting potential
has a periodicity of $\lambda /2$, with a shape depending locally on the
phase $\phi $ between the two beams. It is illustrated through the
theoretical plots of Figs. \ref{fig:pot}b and \ref{fig:pot2D} where, for
sake of simplicity, we used the parameters of experimental demonstration:
the hollow and gaussian intensities are respectively $I_{A}=30$ mW and $%
I_{G}=15$ mW (fig. \ref{fig:pot}a), both beams are polarized $\sigma ^{+}$,
with $\Delta /2\pi =250$ GHz. Fig. \ref{fig:pot}b shows the potential
transverse profile at the bottom of the wells. The geometry of the ring
appears clearly, with a confinement of the order of $r/10$ for $U<150E_{r}$,
and a height for the external barrier of $650E_{r}$. Secondary minima
originating in the residual diffraction produced by the mask, appear inside
the main ring, but because of their weak depth, they should not be annoying
in most applications. A better understanding of the potential distribution
can be obtained from Fig. \ref{fig:pot2D}, where $U$ is plotted in gray
scale versus $r$ and $z$. The potential is constituted by an external
barrier, the height of which varies with $z$. The minimum height $%
U_{A}=650\,E_{r}$,\ in point A of Fig. \ref{fig:pot2D}, occurs at the same $%
z $ as the zero (Fig. \ref{fig:pot}b). The internal barrier has a channel
structure, with the lowest pass in point B (Fig. \ref{fig:pot2D}), at a
height of $U_{B}=110\,E_{r}$, between two successive longitudinal sites.
Atoms with low enough energy are confined in a torus with a half-ellipse
cross-section with axes of the order of 0.1 $\mu $m and 10 $\mu $m. The
diameter of the torus may be chosen of several hundreds of $\mu $m, leading
to a large capture volume. For example, with $r=100$ $\mu $m (Fig. \ref%
{fig:pot2D}) and an initial cloud of diameter 1 mm, 6 \% of the initial
atoms are inside the external cylinder. Thus if the lattice is loaded with a
cloud of $10^{7}$ atoms in 1 mm$^{3}$, which are the typical characteristics
obtained from a MOT, $6\times 10^{5}$ atoms are loaded in 2000 sites,
leading to a filling rate much larger than 1. Thus the potential appears to
hold the required properties, i.e. a 2D confinement and a high filling rate.

To test the feasibility of this lattice, we have implemented an experiment
with the characteristics described above. Atoms of Cesium are initially
cooled in a standard MOT with a detuning of $-3\Gamma $ from resonance. At
time $t=-40$ ms, the magnetic field is turned off, while at time $t=-30$ ms,
the detuning is increased to $-5\Gamma $ and the trap beam intensity is
decreased: this sequence allows us to obtain at time $t=0$ a molasse with a
temperature of 15 $\mu $K, corresponding to an energy of $75E_{r}$, and 10$%
^{7}$ atoms in typically 1 mm$^{3}$.

The hollow and gaussian beams are produced by two diode lasers injected by a
single master diode laser in an extended cavity, which ensure the same
frequency for both beams. For this demonstration, the beams are tuned $250$
GHz above the atomic transition, and are $\sigma ^{+}$ polarized, in order
to increase the depth of the potential wells. In these conditions, the power
of the gaussian and hollow beam, which are respectively 15 and 30 mW, are
sufficient to reach the needed potential depth of $100E_{r}$. The gaussian
beam has a minimum waist, at the level of the MOT, of 140 $\mu $m. The
axicon setup is mounted on an optical rail, to guarantee a good stability of
the beam. The two conical lenses, with a vertex angle of 2$%
{{}^\circ}%
$, are separated by 146 mm. With an incident collimated beam waist equal to
645 $\mu $m and a L focal length of 500 mm, we obtain a hollow beam with $%
r\simeq 1$ mm and $\Delta r\simeq 100$ $\mu $m. A telescope located just
before the trap reduces these values to $r\simeq 100$ $\mu $m and $\Delta
r\simeq 10$ $\mu $m. We obtain in the MOT a transverse distribution of the
hollow beam which is in excellent agreement with the theoretical one.

\begin{figure}[tph]
\centerline{\psfig{figure=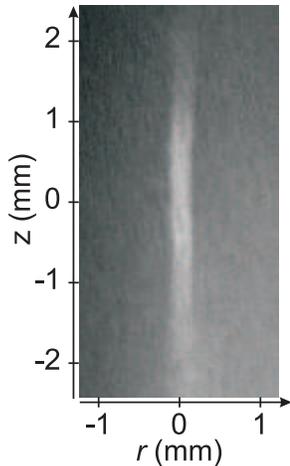,width=1.5in}}
\caption{Snapshot of the atoms in the lattice at time $t=50$ ms.}
\label{fig:picture}
\end{figure}

To load the cold atoms inside the lattice, the later is turned on at a time $%
t<0$, so that when the molasse is switched off, the atoms are already
distributed inside the wells. At time $t=0$, the atoms start to fall under
the effect of gravity, except for those which are trapped in the lattice.
The free atoms needs typically $35$ ms to quit the lattice zone, so that for 
$t>35$ ms, only trapped atoms remain. To observe the atoms, we switch on
during some ms the trap laser beams near resonance, and we used a low-noise
cooled CCD camera to detect the fluorescence emitted by the atoms. A typical
picture is shown in fig. \ref{fig:picture}: as the resolution of the imaging
setup is larger than $\lambda /2$, individual rings cannot be distinguished,
and so atoms appear to be distributed on a cylinder, with a diameter
widening at its ends, due to the defocusing of the beams. The typical number
of atoms obtained at $t=100$ ms is $2\times 10^{4}$, leading to a filling
rate of several atoms by sites.

In conclusion, we propose here a new lattice geometry, namely a 1D stack of
ring traps, enabling large filling rates and high confinement of atoms. A
preliminary experiment allows us to realize such a lattice, loading it
directly from\ a MOT, with a typical filling rate of 10 atoms per site.
Several improvements are necessary before to make such a lattice useful to
e.g. the realization of a quantum gate. In particular, the detuning $\Delta $
should be increased, together with the intensities of lattice beams. The
method to load the cold atoms from\ a MOT in the lattice should be
optimized. To avoid too much losses in the number of atoms, we plan to make
this operation in two steps: first, the hollow beam is switched on just
after the trap beams are switched off, so that the atoms inside the beam are
trapped and remain in the cylinder. Then, the gaussian beam is switched on
progressively, so that the atoms are adiabatically pushed in the ring traps.
This precaution prevents the atoms to be heated by a sudden increasing of
the potential. An interesting alternative would be to load the lattice with
a Bose-Einstein condensate, leading to 1D BEC with a ring geometry.

The Laboratoire de Physique des Lasers, Atomes et Mol\'{e}cules is ``Unit%
\'{e} Mixte de Recherche de l'Universit\'{e} de Lille 1 et du CNRS'' (UMR
8523). The Centre d'Etudes et de Recherches Lasers et Applications (CERLA)
is supported by the Minist\`{e}re charg\'{e} de la Recherche, the R\'{e}gion
Nord-Pas de Calais and the Fonds Europ\'{e}en de D\'{e}veloppement
Economique des R\'{e}gions.


%
%

%
%


\begin{references}
\bibitem{Accelerator} J. Hecker Denschlag, J. E. Simsarian, H. H\"{a}ffner,
C. Mckenzie, A. Browaeys, D. Cho, K. Helmerson, S. L. Rolston and W. D.
Phillips, J. Phys. B: At. Mol. Opt. Phys. {\bf 35}, 3095 (2002)

\bibitem{MottBEC} M. Greiner, O. Mandel, T. Esslinger, T. W. H\"{a}nsch and
I. Bloch, Nature {\bf 415}, 39 (2002)

\bibitem{Expansion} O. Morsch, M. Cristiani, J. H. M\"{u}ller, D. Ciampini
and E. Arimondo, Phys. Rev. A {\bf 66}, 021601(R) (2002)

\bibitem{Excitations} C. Fort, F. S. Cataliotti, L. Fallani, F. Ferlaino, P.
Maddaloni and M. Inguscio, Phys. Rev. Lett. {\bf 90}, 140405 (2003)

\bibitem{1DBEC} K. E. Strecker, G. B. Partridge, A. G. Truscott and R. G.
Hulet, Nature {\bf 417}, 150 (2002); L. Khaykovich, F. Schreck, G. Ferrari,
T. Bourdel, J. Cubizolles, L. D. Carr, Y. Castin and C. Salomon, Science 
{\bf 296}, 1290 (2002)

\bibitem{RingBEC} G. M. Kavoulakis, Phys. Rec. A {\bf 67}, 011601(R) (2003);
R. Kanamoto, H. Saito and M. Ueda, Phys. Rev. A {\bf 67}, 013608 (2003)

\bibitem{QComp2} P. S. Jessen, D. L. Haycock, G. Klose, G. A. Smith, I. H.
Deutsch and G. K. Brennen, QIC {\bf 1}, 20 (2001)

\bibitem{BrightLattice} J. Dalibard, Opt. Commun {\bf 68}, 203 (1988); T.
Walker et al, Phys. Rev. Lett. {\bf 64}, 408 (1990); A. Hemmerich et al,
Europhys. Lett. {\bf 21}, 445 (1993)

\bibitem{Cavity} T. Freegarde and K. Dholakia, Opt. Commun. {\bf 201}, 99
(2002)

\bibitem{DarkLattice} T. Esslinger, F. Sander, A. Hemmerich, T. W. H\"{a}%
nsch, H. Ritsch and M. Weidem\"{u}ller, Opt. Lett. {\bf 21}, 991 (1996)

\bibitem{Axicon} B. D\'{e}pret, Ph. Verkerk, D. Hennequin, Opt. Commun. {\bf %
211}, 31 (2002)

\bibitem{Axicon54} J. H. McLeod, J. Opt. Soc. Am. {\bf 44}, 592 (1954)

\bibitem{BesselGauss} R. M. Herman et al, J. Opt. Soc. Am. A {\bf 8},932
(1991); J. Arlt et al, Opt. commun. {\bf 177}, 297 (2000)

\bibitem{Axicon78} P.-A. B\'{e}langer, Appl. Opt. {\bf 17}, 1080 (1978)

\bibitem{Annular} I. Manek et al, Opt. Commun. {\bf 147}, 67 (1998); L.
Cacciapuoti et al, Eur. Phys. J. D {\bf 14}, 373 (2001)
\end{references}
\end{document}